\documentclass[aps,prl,reprint,floatfix,amsmath,nofootinbib,nobibnotes,preprintnumbers,superscriptaddress]{revtex4-2}

\usepackage[dvipsnames]{xcolor}
\usepackage{color,graphicx,float,xcolor}
\usepackage{amsfonts,amssymb,theorem,mathrsfs,times}
\usepackage{bm}
\usepackage{multirow}
\usepackage{mathtools}
\usepackage{amsfonts,amssymb,theorem,mathrsfs}
\usepackage{dsfont}
\usepackage{setspace}
\usepackage{amsmath}
\usepackage{graphicx}
\usepackage[colorlinks,linkcolor=blue,anchorcolor=blue,citecolor=blue]{hyperref}
\usepackage{ulem}
\usepackage[justification = justified, singlelinecheck = true]{caption}
\usepackage{subcaption}
\usepackage{ragged2e}
\usepackage{tikz}
\usetikzlibrary{patterns,arrows.meta,decorations.pathmorphing,decorations.pathreplacing,calc,positioning,shapes.misc,shapes.geometric}
\usepackage{revsymb}
\usepackage{mathtools}

\begin{document}
\title{Notes on Kerr–Bertotti–Robinson Spacetime}
\preprint{\hfill {\small {ICTS-USTC/PCFT-26-56}}}
\date{\today}

\author{Yu-Sen Zhou}
\email{zhou\_ys@mail.ustc.edu.cn}

\affiliation{Interdisciplinary Center for Theoretical Study and Department of Modern Physics,\\
	University of Science and Technology of China, Hefei, Anhui 230026, China}

\author{Liang-Bi Wu}
\email{wulb@ucas.ac.cn}
\affiliation{School of Fundamental Physics and Mathematical Sciences, Hangzhou Institute for Advanced Study, UCAS, Hangzhou 310024, China}

\author{Ming-Fei Ji}
\email{jimingfei@mail.ustc.edu.cn}

\affiliation{Interdisciplinary Center for Theoretical Study and Department of Modern Physics,\\
	University of Science and Technology of China, Hefei, Anhui 230026, China}

\author{Wen-Tao Fu}
\email{fuwentao2024@mail.ustc.edu.cn}

\affiliation{Interdisciplinary Center for Theoretical Study and Department of Modern Physics,\\
	University of Science and Technology of China, Hefei, Anhui 230026, China}

\author{Li-Ming Cao}
\email{caolm@ustc.edu.cn}
\affiliation{Interdisciplinary Center for Theoretical Study and Department of Modern Physics,\\
	University of Science and Technology of China, Hefei, Anhui 230026, China}
\affiliation{Peng Huanwu Center for Fundamental Theory, Hefei, Anhui 230026, China}

\author{Rong-Gen Cai}
\email{caironggen@nbu.edu.cn}
\affiliation{Institute of Fundamental Physics and Quantum Technology, \& School of Physical Science and Technology, Ningbo University, Ningbo 315211, China}

\begin{abstract}
    The surface $r=\infty$ of the Kerr--Bertotti--Robinson (KBR) spacetime is not the collection of the endpoints of infinitely extended light rays, and the Coulomb type component of gravitational field strength represented by $\Psi_2$ remains nonvanishing there, indicating that this surface is not a real boundary. We construct a natural extension across this surface, which connects the exterior of one KBR region to the interior of a neighboring one and, upon iteration, produces an infinite chain of regions connected by wormhole-like bridges. The extension also exposes the neighboring ring singularity without an intervening horizon, challenging the weak cosmic censorship conjecture and raising the question of whether the extended geometry is stable under perturbations. We therefore study the quasinormal modes (QNMs) of a test massless scalar field on a two-universe scattering segment. For axisymmetric perturbations, the existence of purely imaginary unstable QNMs is analytically proved for every $\ell$, with their origin tied to the chronology-violating region. In the $(\ell,m)=(2,2)$ sector, unstable QNM branches driven by a black-hole-bomb mechanism  are found. Finally, the wormhole geometry produces a double-barrier cavity and families of weakly damped QNMs, suggesting echo-like responses.
\end{abstract}

\maketitle
\section{Introduction}
Astrophysical black holes are often immersed in strong electromagnetic fields. The recently constructed Kerr--Bertotti--Robinson (KBR) solution~\cite{Podolsky:2025tle}, commonly interpreted as a rotating black hole in an asymptotically uniform electromagnetic field, has motivated extensive studies of its phenomenology and exact structure~\cite{Ovcharenko:2025cpm,Astorino:2025lih,Zeng:2025olq,Wang:2025vsx,Zeng:2025tji,Wang:2025bjf,Ali:2025beh,Zhang:2025ole,Liu:2025wwq,Li:2025rtf,Mirkhaydarov:2026fyn,Wang:2026czl,Mustafa:2026gly,Ahmed:2025ril,Siahaan:2025ngu,Gray:2025lwy,Siahaan:2026tuf,Lu:2026kcm,Hu:2026slp,Rehman:2026rzq,Wan:2026lca,Roy:2026poj,Hassanabadi:2026fzz,Huang:2026qzj,Ahmed:2026ozq,Singh:2026rbz}. Despite this rapid progress, the global interpretation of the KBR spacetime remains unclear, largely because the nature of the surface $r=\infty$ is not well understood, where $r$ is the radial coordinate in the Boyer–Lindquist-type coordinates. Although it is known not to represent null infinity~\cite{Podolsky:2025tle}, its continuation and global meaning are still unexplored. Light rays and timelike observers can readily reach this surface, yet the spacetime beyond it remains unknown. 

To resolve this issue, we construct a smooth extension across it. The extension identifies $r=+\infty$ of one KBR region with $r=-\infty$ of a neighboring one and, upon iteration, produces an infinite chain in which each exterior forms a wormhole-like bridge to the interior of the next universe. Thus, rather than revealing a conventional null infinity, the extension exposes a larger connected spacetime for which the usual global black-hole interpretation does not apply.

The extension also exposes the ring singularity of the neighboring KBR region, with no horizon separating it from the original exterior, thereby violating  the weak cosmic censorship conjecture (WCCC). The standard modern formulation of the WCCC, however, concerns the evolution of generic initial data and assumes an appropriate complete future null infinity~\cite{Wald:1997wa}, which is absent here. This raises the possibility that the extended KBR spacetime, if realizable through evolution at all, may require highly special initial data. An instability under perturbations would support this interpretation and would be consistent with the spirit of the WCCC. We are therefore led to examine the stability of the extended geometry.

The stability of this extended geometry is probed through its quasinormal modes (QNMs)~\cite{Kokkotas:1999bd,Nollert:1999ji,Berti:2009kk,Konoplya:2011qq,Berti:2025hly}, the characteristic resonances of linear perturbations whose imaginary frequencies determine whether perturbations decay or grow. A test massless scalar field is considered on a two-universe scattering region extending from the outer horizon of one KBR region to the inner horizon of the neighboring one. By slightly modifying the method of Ref.~\cite{Dotti:2011ix}, we analytically establish the existence of purely imaginary, exponentially growing $m=0$ QNMs for every $\ell$, and verify them numerically. As in Kerr~\cite{Dotti:2011ix}, the associated radial confining region coincides with the equatorial region containing closed timelike curves, suggesting that the instability is closely related to this causal pathology. Since the scattering region is not globally hyperbolic, however, the existence of these growing modes does not by itself establish a conventional dynamical instability arising from generic Cauchy data.

We next turn to the nonaxisymmetric quadrupolar sector, $(\ell,m)=(2,2)$, both to test whether the growing spectrum is peculiar to the special $m=0$ construction and to probe rotational effects such as superradiance. We find unstable QNM branches for sufficiently rapid rotation and sufficiently weak electromagnetic fields. At the onset of instability, these branches pass through purely real modes for which absorption at the outer horizon of one universe is exactly balanced by superradiant extraction from the inner horizon of the neighboring one. The unstable branches emerge continuously from these marginal modes, providing strong evidence for a superradiant origin. Thus, the unstable modes persist in a nonaxisymmetric sector more directly connected with rotational and radiative dynamics.

Although the extended KBR spacetime is not a black hole in the standard global sense, its outer Killing horizon admits a quasi-local interpretation~\cite{Ashtekar:1998sp,ashtekar:2004cn}, and hence providing a natural setting for comparison with Kerr. Even in the weak-electromagnetic-field regime, however, its QNM spectrum remains distinct from that of Kerr because the scattering region extends through the wormhole bridge into the neighboring universe rather than terminating at null infinity. This wormhole geometry also leaves a characteristic spectral imprint. The effective potential typically forms a cavity between two barriers across the gluing surface, accompanied by families of weakly damped QNMs characteristic of echo modes. Their presence suggests echo-like responses associated with the wormhole structure.

\section{The Natural Extension and Global Structure}\label{sec:setup}
The line element of the Kerr--Bertotti--Robinson (KBR)~\cite{Podolsky:2025tle} solution reads:
\begin{eqnarray}\label{ds}
	\mathrm{d}s^{2}&=&
	\frac{1}{\Omega^{2}}\left[-\frac{Q}{\rho^{2}}
	\left(\mathrm{d}t-a\Delta_xC\mathrm{d}\varphi\right)^{2}
	+\frac{\rho^{2}}{Q}\mathrm{d}r^{2}
	\right.\nonumber\\
	&&\left.
	+\frac{\rho^{2}}{P\Delta_x}\mathrm{d}x^{2}
	+\frac{P\Delta_x}{\rho^{2}}
	\left(a\mathrm{d}t-\rho_0^2C\mathrm{d}\varphi\right)^{2}
	\right]\, ,
\end{eqnarray}
where the metric functions are
\begin{eqnarray}
	\rho_0^2(r)&=& r^{2}+a^{2}\, ,\nonumber\\
	\rho^{2}(r,x) &=& r^{2}+a^{2}x^{2}\, ,\nonumber\\
	P(x) &=& 1+\left(\frac{1}{C}-1\right)x^{2}\, ,\nonumber\\
	Q(r) &=& (1+B^{2}r^{2})\Delta\, ,\nonumber\\
	\Omega^{2}(r,x) &=& (1+B^{2}r^{2})-B^{2}\Delta x^{2}\, ,\nonumber\\
	\Delta(r) &=& A r^{2}-2\mu r+a^{2}\, ,\nonumber\\
	\Delta_x(x)&=&1-x^2\, ,
\end{eqnarray}
and
\begin{eqnarray}
	I_{1}&=&1-\frac{1}{2}B^{2}a^{2}\,,\quad
	I_{2}=1-B^{2}a^{2}\, ,\quad
	\mu=M\frac{I_2}{I_1}\, ,\nonumber\\
	A&=&1-B^{2}M^{2}\frac{I_{2}}{I_{1}^{2}}\, ,\quad
	C=\left[1+B^{2}\left(\frac{M^{2}I_{2}}{I_{1}^{2}}-a^{2}\right)\right]^{-1}\, .
\end{eqnarray}
Here $M$, $a$, and $B$ denote the mass parameter, rotation parameter, and strength of the external uniform magnetic field, respectively. We have switched from the usual polar angle $\theta$ to the variable $x=\cos\theta$, and explicitly introduced the conicity factor $C$ to remove conical defects on the symmetry axis, so that $\varphi\in[0,2\pi)$~\cite{Destounis:2020pjk, Xiong:2023usm, Chen:2024rov}.

The complex electromagnetic potential is given by
\begin{eqnarray}
	A_\mu \mathrm{d}x^\mu
	&=&\frac{\mathrm{e}^{\mathrm{i}\nu}}{2B}\Bigg[
		\Omega_{,r}\frac{a\mathrm{d}t-\rho_0^2\mathrm{d}\varphi}{r+\mathrm{i}ax}\nonumber\\
		&&-\mathrm{i}\Omega_{,x}\frac{\mathrm{d}t-a\Delta_x\mathrm{d}\varphi}{r+\mathrm{i} a x}+(\Omega-1)\mathrm{d}\varphi
		\Bigg]\,.
\end{eqnarray}
Here $\nu$ is the duality-rotation parameter, and the physical gauge potential is obtained as $A_\mu^{\text{real}}=2\mathrm{Re}(A_\mu)$. The corresponding field strength $F=\mathrm{d}A^{\text{real}}$, together with the metric~(\ref{ds}), satisfies the Einstein-Maxwell equations~\cite{Podolsky:2025tle}.

In the following, we restrict to the subextremal $A>0$ case, in which $Q(r)$ admits two distinct positive real roots, denoted by $r_{\mathrm{o}}$ and $r_{\mathrm{i}}$. We refer to the corresponding Killing horizons as the outer and inner horizons, respectively. For convenience in the subsequent analysis, we denote the pair of complex-conjugate roots of $Q(r)$ by $r_{+}=\mathrm{i}/B$ and $r_{-}=-\mathrm{i}/B$.

Consider a radially outgoing null geodesic, affinely parameterized by $s$, satisfying $\dot t-\dot r_*=\dot x=\dot{\varphi}-\dot{\psi}=0$, where a dot denotes differentiation with respect to $s$. Here, the tortoise coordinate $r_*$ and the dragged azimuthal coordinate $\psi$ are defined by $\mathrm{d}r_*=\rho_0^2\mathrm{d}r/Q,\mathrm{d}\psi=a\mathrm{d}r_*/(\rho_0^2C)$. The resulting tortoise coordinate tends to a constant as $r\to\infty$, which is fixed as $0$ by an appropriate choice of the integration constant. The affine parameter of the light satisfies $\mathrm{d}s/\mathrm{d}r=(E\Omega^2)^{-1}=\mathcal{O}(r^{-2})$, where $E$ denotes the conserved energy of the light ray. Thus, the outgoing null ray reaches $r=\infty$ within a finite affine parameter. The facts that light can reach $r=\infty$ within finite affine parameter and that $\Psi_2$ approaches a nonzero constant~\cite{Podolsky:2025tle} as $r\to\infty$ both indicate that this surface is not a genuine physical boundary. Indeed, timelike observers can also reach and cross $r=\infty$ from the vicinity of the Killing horizon in finite proper time, with the crossing itself being entirely regular and geodesic. These observations naturally raise the question of how the spacetime should be extended across this surface.

To construct such an extension, we introduce a new radial coordinate $y$ that remains regular at $r=\infty$. Requiring the metric component $g_{yy}$ to be regular gives $\mathrm{d}y/\mathrm{d}r\sim r^{-2}+O(r^{-3})$, so any such radial coordinate is proportional to $1/r$ at leading order, up to a smooth reparametrization. Specifically, we choose $y=-r_{\text{o}}/r$. In terms of $(t,y,x,\varphi)$, the metric~\eqref{ds} becomes
\begin{eqnarray}
	\mathrm{d}s^2&=&\frac{1}{\widehat{\Omega}^2}\left[
	-\frac{\widehat{Q}}{\widehat{\rho}^2}
	\left(\mathrm{d}t-a\Delta_xC\mathrm{d}\varphi\right)^2
	+\frac{\widehat{\rho}^2}{\widehat{Q}}r_{\text{o}}^2\mathrm{d}y^2
	\right.\nonumber\\
	&&\left.
	+\frac{\widehat{\rho}^2}{P\Delta_x}\mathrm{d}x^2
	+\frac{P\Delta_x}{\widehat{\rho}^2}
	\left(ay^2\mathrm{d}t-\widehat{\rho}_0^2C\mathrm{d}\varphi\right)^2
	\right]\,,
\end{eqnarray}
where
\begin{eqnarray}
	\widehat{\rho}_0^2&=&y^2\rho_0^2(-r_{\text{o}}/y)=r_{\text{o}}^2+a^2y^2\,,\nonumber\\
	\widehat{\rho}^2&=&y^2\rho^2(-r_{\text{o}}/y,x)=r_{\text{o}}^2+a^2x^2y^2\,,\nonumber\\
	\widehat{Q}&=&y^4Q(-r_{\text{o}}/y)=(y^2+B^2r_{\text{o}}^2)\widehat{\Delta}\,,\nonumber\\
	\widehat{\Delta}&=&y^2\Delta(-r_{\text{o}}/y)=Ar_{\text{o}}^2+2\mu r_{\text{o}}y+a^2y^2\,,\nonumber\\
	\widehat{\Omega}&=&\lvert y\rvert\Omega(-r_{\mathrm{o}}/y,x)=\sqrt{(y^2+B^2r_{\mathrm{o}}^2)-B^2\widehat{\Delta}x^2}\,.\nonumber
\end{eqnarray}
These hatted quantities are smooth at $y=0$, and hence the metric extends smoothly across this surface. Here we choose $\Omega$ to have the same sign in the two KBR universes, so smooth continuation of the electromagnetic field requires $\nu\to\nu+\pi$. Equivalently, one may keep $\nu$ fixed and let $\Omega$ change sign across $y=0$. The geometry is unchanged since the metric depends only on $\Omega^2$. The hypersurface $y=0$ is timelike, and the induced metric and extrinsic curvature agree on the two sides, so no matter shell is introduced there~\cite{Israel:1966rt, Mars:1993mj}. Since the metric and Maxwell field are analytic in $y$ at $y=0$, their continuation across this surface is locally unique up to regular coordinate reparametrizations and electromagnetic gauge transformations. In this sense, the extension is natural rather than an arbitrary gluing.

We may therefore continue the spacetime across $y=0$. In the extended region, we denote the radial coordinate by $\bar{r}=-r_{\text{o}}/y$. In this way, the surface $r=+\infty$ of the first universe is smoothly connected to the surface $\bar r=-\infty$ of a neighboring KBR universe with the same metric parameters. Since each continuation shifts $\nu$ by $\pi$, adjacent universes cannot be identified. A period-two identification would be possible after two continuations, but we keep the universes distinct. The resulting geometry has a wormhole-like structure, with $y=0$ connecting the exterior of one KBR universe to the interior of the neighboring one. Remarkably, the bridge is supported entirely by the electromagnetic field and violates none of the standard energy conditions. 

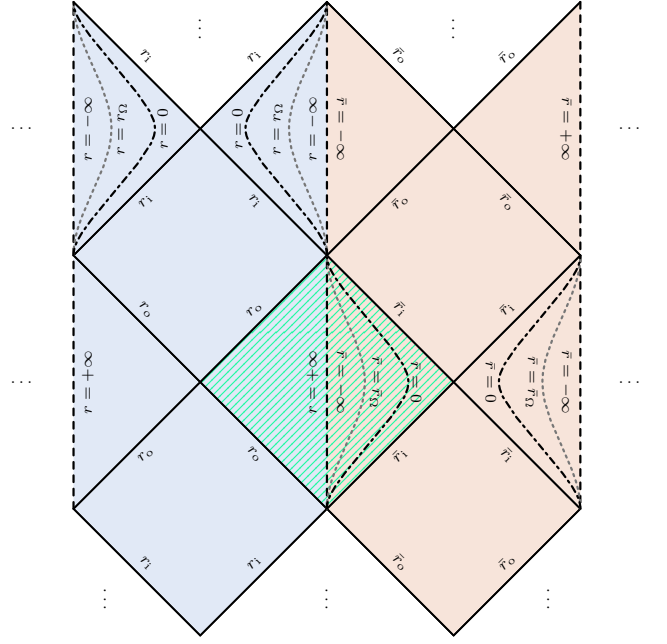
\begin{figure}[htbp]
	\centering
	\resizebox{\columnwidth}{!}{
	\begin{tikzpicture}[scale=0.2, line width=1pt, line join=round,
  line cap=round, font=\scriptsize]

  \coordinate (A10) at (10,0);  \coordinate (A01) at (0,10);
  \coordinate (A21) at (20,10); \coordinate (A12) at (10,20);
  \coordinate (A03) at (0,30);  \coordinate (A23) at (20,30);
  \coordinate (A14) at (10,40); \coordinate (A05) at (0,50);
  \coordinate (A25) at (20,50); \coordinate (A30) at (30,0);
  \coordinate (A41) at (40,10); \coordinate (A32) at (30,20);
  \coordinate (A43) at (40,30); \coordinate (A34) at (30,40);
  \coordinate (A45) at (40,50);

  \fill[RoyalBlue!10] (A05)--(A01)--(A10)--(A21)--(A25)--(A14)--cycle;
  \fill[BrickRed!10] (A25)--(A21)--(A30)--(A41)--(A45)--(A34)--cycle;
  \fill[pattern=north east lines,pattern color=TealBlue]
    (A21)--(A32)--(A23)--(A12)--cycle;

  \foreach \p/\q in {
    A10/A01,A10/A21,A01/A12,A21/A12,A12/A03,A12/A23,
    A03/A14,A23/A14,A14/A05,A14/A25,A30/A21,A30/A41,
    A21/A32,A41/A32,A32/A23,A32/A43,A23/A34,A43/A34,
    A34/A25,A34/A45}{\draw[thick] (\p)--(\q);}

  \draw[thick] (A10)--(A01) node[midway,sloped,above] {$r_{\mathrm{i}}$};
  \draw[thick] (A10)--(A21) node[midway,sloped,above] {$r_{\mathrm{i}}$};
  \draw[thick] (A01)--(A12) node[midway,sloped,below] {$r_{\mathrm{o}}$};
  \draw[thick] (A21)--(A12) node[midway,sloped,below] {$r_{\mathrm{o}}$};
  \draw[thick] (A12)--(A03) node[midway,sloped,above] {$r_{\mathrm{o}}$};
  \draw[thick] (A12)--(A23) node[midway,sloped,above] {$r_{\mathrm{o}}$};
  \draw[thick] (A03)--(A14) node[midway,sloped,below] {$r_{\mathrm{i}}$};
  \draw[thick] (A23)--(A14) node[midway,sloped,below] {$r_{\mathrm{i}}$};
  \draw[thick] (A14)--(A05) node[midway,sloped,above] {$r_{\mathrm{i}}$};
  \draw[thick] (A14)--(A25) node[midway,sloped,above] {$r_{\mathrm{i}}$};
  \draw[thick] (A30)--(A21) node[midway,sloped,above] {$\bar r_{\mathrm{o}}$};
  \draw[thick] (A30)--(A41) node[midway,sloped,above] {$\bar r_{\mathrm{o}}$};
  \draw[thick] (A21)--(A32) node[midway,sloped,below] {$\bar r_{\mathrm{i}}$};
  \draw[thick] (A41)--(A32) node[midway,sloped,below] {$\bar r_{\mathrm{i}}$};
  \draw[thick] (A32)--(A23) node[midway,sloped,above] {$\bar r_{\mathrm{i}}$};
  \draw[thick] (A32)--(A43) node[midway,sloped,above] {$\bar r_{\mathrm{i}}$};
  \draw[thick] (A23)--(A34) node[midway,sloped,below] {$\bar r_{\mathrm{o}}$};
  \draw[thick] (A43)--(A34) node[midway,sloped,below] {$\bar r_{\mathrm{o}}$};
  \draw[thick] (A34)--(A25) node[midway,sloped,above] {$\bar r_{\mathrm{o}}$};
  \draw[thick] (A34)--(A45) node[midway,sloped,above] {$\bar r_{\mathrm{o}}$};

  \draw[dashed] (0,10)--(0,30) node[midway,sloped,below] {$r=+\infty$};
  \draw[dashed] (0,30)--(0,50) node[midway,sloped,below] {$r=-\infty$};
  \draw[dashed] (20,10)--(20,30) node[midway,sloped,above] {$r=+\infty$};
  \draw[dashed] (20,30)--(20,50) node[midway,sloped,above] {$r=-\infty$};
  \path (20,50)--node[midway,sloped,above] {$\bar r=-\infty$}(20,30);
  \path (20,30)--node[midway,sloped,above] {$\bar r=-\infty$}(20,10);
  \draw[dashed] (40,30)--(40,10) node[midway,sloped,below] {$\bar r=-\infty$};
  \draw[dashed] (40,50)--(40,30) node[midway,sloped,below] {$\bar r=+\infty$};

  \def\rhoOmega{0.5082406860240279}
  \def\rhoZero{1.5907142458953558}
  \def\qmin{-1}
  \def\qmax{1}
  \tikzset{
    omega/.style={draw={rgb,255:red,119;green,119;blue,119},
      line width=1pt,dotted},
    zero/.style={draw=black,line width=1pt,dash dot},
    curve label/.style={font=\scriptsize,inner sep=0pt}
  }
  \newcommand{\exactcurve}[5]{%
    \draw[#1] ({20*(#3)},{20*(#5-0.5)})
      plot[domain=\qmin:\qmax,samples=241,variable=\q]
      ({20*(#3 + #4*(atan(10000*\q*\q*\q+#2)
                    -atan(10000*\q*\q*\q-#2))/360)},
       {20*(#5 + (atan(10000*\q*\q*\q-#2)
                    +atan(10000*\q*\q*\q+#2))/360)})
      -- ({20*(#3)},{20*(#5+0.5)});
  }

  \exactcurve{omega}{\rhoOmega}{1}{1}{1}
  \exactcurve{zero}{\rhoZero}{1}{1}{1}
  \exactcurve{omega}{\rhoOmega}{0}{1}{2}
  \exactcurve{zero}{\rhoZero}{0}{1}{2}
  \exactcurve{omega}{\rhoOmega}{1}{-1}{2}
  \exactcurve{zero}{\rhoZero}{1}{-1}{2}
  \exactcurve{omega}{\rhoOmega}{2}{-1}{1}
  \exactcurve{zero}{\rhoZero}{2}{-1}{1}

  \node[curve label,rotate=90]  at (3.85,40) {$r=r_\Omega$};
  \node[curve label,rotate=90]  at (7.05,40) {$r=0$};
  \node[curve label,rotate=90]  at (16.15,40) {$r=r_\Omega$};
  \node[curve label,rotate=90]  at (12.95,40) {$r=0$};
  \node[curve label,rotate=270]  at (23.85,20) {$\bar r=\bar r_\Omega$};
  \node[curve label,rotate=270]  at (27.05,20) {$\bar r=0$};
  \node[curve label,rotate=270] at (36.15,20) {$\bar r=\bar r_\Omega$};
  \node[curve label,rotate=270] at (32.95,20) {$\bar r=0$};

  \node at (-4,20) {$\cdots$};
  \node at (-4,40) {$\cdots$};
  \node at (44,20) {$\cdots$};
  \node at (44,40) {$\cdots$};
  \node[rotate=90,anchor=center] at (10,48) {$\cdots$};
  \node[rotate=90,anchor=center] at (30,48) {$\cdots$};
  \node[rotate=90,anchor=center] at (2.5,3) {$\cdots$};
  \node[rotate=90,anchor=center] at (20,3) {$\cdots$};
  \node[rotate=90,anchor=center] at (37.5,3) {$\cdots$};
\end{tikzpicture}}
	\caption{\justifying Penrose diagram of the extended spacetime considered in this work. Only a local two-universe segment is displayed here, as the full diagram is obtained by infinitely repeating this minimal building block. The left and right colored regions correspond to the first and second universes, respectively, while the ellipses indicate the infinite continuation of the diagram by further gluing. The straight dashed lines denote the timelike surfaces $r=\pm\infty$ and $\bar r=\pm\infty$, across which neighboring universes are connected. The dash-dotted curves mark $r=0$ and $\bar r=0$, with ring singularities at $x=0$. The gray dotted curves indicate $r_\Omega=\bar r_\Omega=-\mu/(1-A)$, where $\Omega$ vanishes at $x=\pm1$. The hatched diamond denotes the scattering region considered here.}
	\label{fig:kbr-penrose}
\end{figure}

\begin{figure}
	\includegraphics[width=\columnwidth]{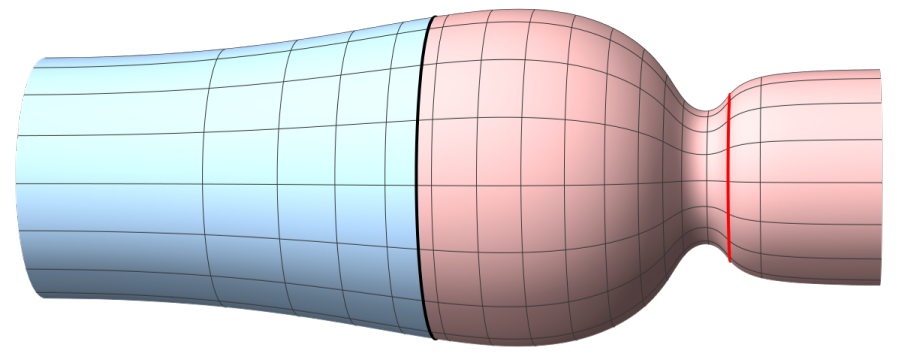}
	\caption{\justifying 
    Isometric embedding of the fixed-$t$, fixed-$x=x_{0}=0.7$ spatial slice for $M=1$, $a/M=0.9$, and $BM=0.6$. The black and red circles mark the gluing surface $r=+\infty\sim\bar r=-\infty$ and $\bar r=0$, respectively. We choose $\lvert x_{0}\rvert$ sufficiently large to avoid  the ring singularity, the chronology-violating region, and the region where Euclidean embedding fails because the circumferential radius varies faster than the proper radial distance. The same obstructions occur in the negative-$r$ extension of Kerr. The compact coordinate $\chi=\arctan(r/r_{\mathrm{o}})$ is chosen to obtain this standard axisymmetric isometric embedding.}
	\label{fig:embed}
\end{figure}

Following the same null ray beyond $y=0$, its affine parameter satisfies $\mathrm{d}s/\mathrm{d}y=r_{\mathrm{o}}/(E\widehat{\Omega}^{2})=\mathcal{O}(y^{-2})$
as $y\to+\infty$. Provided that the ray avoids the ring singularity and the isolated points ${\bar r_\Omega=-\mu/(1-A),x=\pm1}$ where $\Omega=0$, it can pass through $\bar r=0$ and reach $\bar r=+\infty$ within finite affine parameter. The same construction can therefore be iterated, producing an infinite chain of KBR universes. The extension also exposes the ring singularity at $\bar r=0$, $x=0$ in the neighboring universe, with no horizon separating it from the original exterior. Thus, the extended spacetime violates the WCCC. A schematic Penrose diagram of the extended spacetime is shown in Fig.~\ref{fig:kbr-penrose}, together with an isometric embedding of a fixed-$t$, fixed-$x$ spatial slice in Fig.~\ref{fig:embed}.

\section{Wave Scattering on the Extended Spacetime}\label{sec:scattering}
Motivated by the apparent violation of WCCC, we focus on a single scattering segment and study the scalar QNM spectrum as a probe of the spectral stability of the extended spacetime. Specifically, we consider a test massless Klein--Gordon field satisfying $\square\Psi=0$. With the ansatz
\begin{eqnarray}
	\Psi(t,r,x,\varphi)=\Omega\mathrm{e}^{-\mathrm{i}(\omega t-m\varphi)}\frac{\Psi^{(r)}(r)}{r}\Psi^{(x)}(x)\, ,
\end{eqnarray}
the equation separates into radial and angular parts:
\begin{eqnarray}
	\frac{\mathrm{d}}{\mathrm{d}r}\left(\frac{Q}{r^2}
	\frac{\mathrm{d}\Psi^{(r)}}{\mathrm{d}r}\right)
	+\frac{1}{r^2}\Bigg[\frac{\left(\omega\rho_0^2-a m/C\right)^2}{Q}\nonumber\\
		-\frac{2}{r^2}(\mu r-a^2)-\lambda\Bigg]\Psi^{(r)}=0\,,\label{eqr}\\
	\frac{\mathrm{d}}{\mathrm{d}x}\left(P\Delta_x
	\frac{\mathrm{d}\Psi^{(x)}}{\mathrm{d}x}\right)
	-\Bigg[\frac{\left(a\omega\Delta_x-m/C\right)^2}{P\Delta_x}\nonumber\\
		+2(P-1)-\lambda\Bigg]\Psi^{(x)}=0\,.\label{eqx}
\end{eqnarray}
Here $\lambda$ is the separation constant, reducing to the Kerr value $\lambdabar=A_{\ell m}(a\omega)+a^2\omega^2-2am\omega$ as $B\to0$, while $m\in\mathbb{Z}$ since the conicity factor $C$ has already been incorporated into the metric. Despite the explicit $r^{-1}$ factor in the ansatz, the field is regular at $r=0$, since Eq.~\eqref{eqr} gives $\Psi^{(r)}=\mathcal{O}(r)$ and hence $\Psi^{(r)}/r$ remains finite. A Liouville transformation can be applied to cast the Sturm-Liouville equation~\eqref{eqr} into a Schr\"odinger-like form:
\begin{eqnarray}
	&&\frac{\mathrm{d}^2\Psi^{(r_*)}}{\mathrm{d}r_*^2}+V\Psi^{(r_*)}=0\,,\label{eqrstar}\\
	&&V=\frac{Q}{\rho_0^4}\left[\frac{\left(\omega\rho_0^2-a m/C\right)^2}{Q}
		+a^2r\rho_0\frac{\mathrm{d}}{\mathrm{d}r}\left(\frac{Q}{r^2\rho_0^3}\right)\right.\nonumber\\
		&&\left.\hspace{1.6cm}
		-\frac{2}{r^2}(\mu r-a^2)-\lambda\right]\,.\nonumber
\end{eqnarray}
We take the scattering domain to extend from the outer horizon $r=r_{\mathrm{o}}$ of the first universe to the inner horizon $\bar r=\bar r_{\mathrm{i}}$ of the neighboring universe. The tortoise coordinate $r_*$, normalized by $r_*=0$ at $y=0$, covers the entire segment, with $r_*\to-\infty$ as $r\to r_{\mathrm{o}}$ and $r_*\to+\infty$ as $\bar r\to\bar r_{\mathrm{i}}$.

\begin{figure}[htbp]
	\centering
	\includegraphics[width=0.4\textwidth]{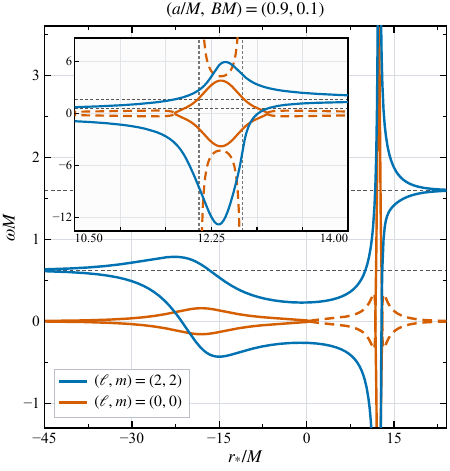}
	\caption{\justifying
	The roots $\omega_i(r_*)$ that satisfy $V(r_*;\omega_i,\lambda(\omega_i))=0$. Solid curves denote real roots, while dashed curves denote the imaginary parts of the purely imaginary roots. The inset enlarges the horizontal scale while extending the vertical range. The horizontal dashed lines indicate the asymptotic frequencies $m\varpi_{\mathrm{o}}$ and $m\varpi_{\mathrm{i}}$, while the vertical dashed lines mark the equatorial locations where $g_{\varphi\varphi}=0$. Here we have chosen $M=1$, $a/M=0.9$, $BM=0.1$ for visualization.}
	\label{fig:po}
\end{figure}

To characterize the effective potential, following~\cite{Comins, Cardoso:2007az, Patrick:2018orp}, we consider the WKB regime, where the radial solution takes the form
\begin{eqnarray}
    \Psi^{(r_*)} \sim A_0 \exp\left[\mathrm{i}\int k(r_*;\omega,\lambda)\mathrm{d}r_*\right]\, ,
\end{eqnarray}
with $A_0$ a slowly varying amplitude and the local wave number $k$ satisfying $k^2=V$. We then plot in Fig. \ref{fig:po} the real-frequency roots $\omega_i(r_*)$ defined by $V(r_*;\omega_i,\lambda(\omega_i))=0$. Each $\omega_i(r_*)$ is the frequency at which $r_*$ becomes a WKB turning point between oscillatory and evanescent regions. For $(\ell,m)=(0,0)$, we also show the imaginary parts of purely imaginary roots as dashed curves; some emerge continuously from the real branches while others form disconnected branches. The roots approach $m\varpi_{\mathrm{o}}$ and $m\bar{\varpi}_{\mathrm{i}}$ as $r_*\to-\infty$ and $r_*\to+\infty$, respectively, where $\varpi_j=\left.a/(C\rho_0^2)\right|_{r=r_j}$ denotes the angular velocity of the corresponding horizon. Typically, the effective potential has two peaks, one in each universe, forming a cavity across the wormhole bridge. Only for sufficiently large $BM$ or small $a/M$ does the peak in the first universe disappear. Thus, the wormhole-like structure is also reflected in the wave dynamics.

We impose an ingoing condition at the outer horizon of the first universe and an outgoing condition at the inner horizon of the neighboring universe,
\begin{eqnarray}
	\Psi^{(r)}&\sim& \mathrm{e}^{-\mathrm{i}(\omega-m\varpi_{\text{o}})r_*}\, ,\quad r\to r_{\text{o}}\, , \nonumber\\
	\bar{\Psi}^{(\bar{r})}&\sim& \mathrm{e}^{\mathrm{i}(\omega-m\bar{\varpi}_{\text{i}})\bar{r}_*}\, ,\quad\;\;\; \bar{r}\to \bar{r}_{\text{i}}\, .
\end{eqnarray}
The latter describes waves leaving the scattering region toward the white-hole-like region beyond the inner horizon of the second universe.

As in Kerr, the extended KBR spacetime contains a region in the negative-$r$ sector where $g_{\varphi\varphi}<0$, so that the axial Killing orbits are closed timelike curves (CTCs). In the axisymmetric $m=0$ sector, the scalar wave equation becomes elliptic upon entering this region. Motivated by the suggested connection between such a CTC region and unstable modes in Kerr~\cite{Dotti:2011ix}, we find that the method of~\cite{Dotti:2011ix}, with minor modifications, can be applied to the KBR spacetime. Specifically, it establishes the existence of purely imaginary, exponentially growing $m=0$ modes for every $\ell$. We also verified these unstable modes numerically.

Figure~\ref{fig:po} provides a complementary physical picture of these unstable modes. The purely imaginary root branches diverge precisely where $g_{\varphi\varphi}=0$ on the equatorial plane, so that the radial confining region between them coincides with the equatorial CTC region. Within this region the mode is spatially oscillatory, while outside it decays exponentially toward both boundaries. This bound-state-like configuration can support a localized QNM with $\operatorname{Im}\omega>0$, strongly suggesting that the instability is tied to the chronology-violating region.

However, since the scattering region contains CTCs, it is not globally hyperbolic and does not admit a standard globally well-posed Cauchy evolution. Therefore, the growing $m=0$ modes alone do not establish a dynamical instability from generic initial data. We therefore further examine the $m=2$ sector, which describes nonaxisymmetric quadrupolar perturbations and is more directly connected with the dynamics relevant to gravitational radiation.

For $m=2$ and generic complex $\omega$, both the radial and angular equations can be reduced to Heun form. We construct local Heun solutions satisfying the appropriate boundary conditions and determine the QNMs by matching the radial solutions across the two coordinate patches together with the angular regularity condition~\cite{Suzuki:1998vy, Fiziev:2011mm, Hatsuda:2020sbn, Motohashi:2021zyv, Wu:2025wbp}.

\section{Quasinormal Mode Spectrum}
\begin{figure}[htbp]
    \centering
    \includegraphics[width=\columnwidth]{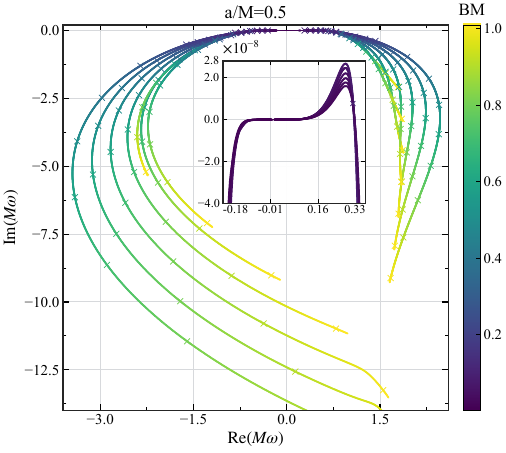}
    \includegraphics[width=\columnwidth]{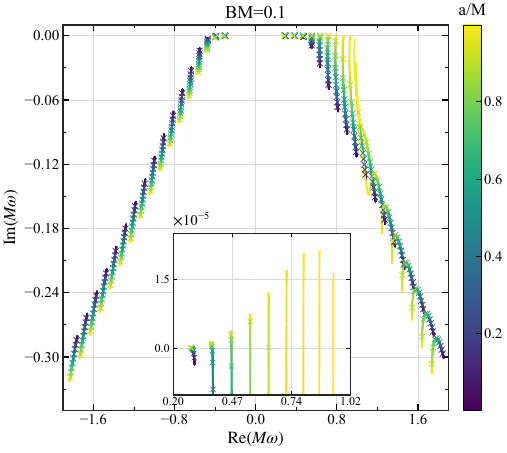}
    \caption{\justifying
    The $\ell=2$, $m=2$ QNM spectrum. In the upper panel, $BM$ is varied at fixed $a/M=0.5$, while in the lower panel, $a/M$ is varied at fixed $BM=0.1$. The color indicates the value of the varied parameter, and small crosses mark $0.05,0.10,0.15,\ldots$. The insets magnify the vicinity of the real axis, where some branches cross into $\operatorname{Im}(M\omega)>0$.}
    \label{fig:qnm-spectrum}
\end{figure}

We now present the QNM spectrum in the $\ell=2$, $m=2$ sector. Figure~\ref{fig:qnm-spectrum} shows two representative scans, varying $BM$ at fixed $a/M=0.5$ and varying $a/M$ at fixed $BM=0.1$. In each case, the varied parameter is scanned from $0.001$ to $0.001$ below its maximal value allowed by subextremality and $A>0$. Several QNM branches cross into the upper half of the complex-frequency plane for sufficiently large $a/M$ and sufficiently small $BM$.

The instability is reminiscent of the black-hole bomb~\cite{Press:1972zz, Cardoso:2004nk}. At each marginal real mode, conservation of the radial Wronskian shows,
\begin{eqnarray}
    (\omega-m\varpi_{\mathrm{o}})|A_{\mathrm{o}}|^{2}+(\omega-m\bar{\varpi}_{\mathrm{i}})|A_{\mathrm{i}}|^{2}=0\,,
\end{eqnarray}
where $A_{\mathrm{o}}$ and $A_{\mathrm{i}}$ denote the amplitudes at the outer and inner horizons, respectively. This requires an exact balance of the $\partial_t$ Killing-energy fluxes through the two horizons. Since $m\varpi_{\mathrm{o}}<m\bar{\varpi}_{\mathrm{i}}$, the frequency $\omega$ must lie between these two frequencies. The energy absorbed by the outer horizon is exactly balanced by the rotational energy extracted from the neighboring inner horizon through superradiance. Superradiant amplification, however, requires a continuous supply of incident flux, for which the intermediate potential cavity, visible in the effective potential in Fig. \ref{fig:po},  provides a natural feedback mechanism. Trapped modes can repeatedly leak through the barrier toward the inner horizon, couple to the negative-Killing-energy branch, and return an amplified reflected component to the cavity. The unstable branches emerge continuously from these marginal modes, suggesting that this balance is tipped toward net superradiant amplification. Moreover, such a cavity appears in a broader range of parameters than the unstable modes, which is consistent with this interpretation.

Although the extended KBR spacetime is not a black hole in the standard global sense, its outer Killing horizon permits a quasi-local comparison with Kerr. Even for weak $BM$, however, the scattering region extends from the outer horizon of one universe to the inner horizon of the neighboring one rather than to null infinity. The QNMs therefore remain distinct from the Kerr spectrum.

The wormhole structure is also reflected in the QNM spectrum. For fixed $a/M=0.5$ and varying $BM$, the spectrum contains many weakly damped modes whose trajectories lie close to and approximately parallel to the real axis. Such modes are characteristic of echo modes associated with a potential cavity. Together with the two potential barriers across the gluing surface as shown in Fig. \ref{fig:po}, this spectral structure suggests that waves may undergo repeated reflections between the barriers and produce echo-like signals.

\section{Conclusion and Discussion}
In this paper, an explicit smooth extension of the KBR spacetime is constructed. This extension arises naturally from the analytic continuation of the geometry and electromagnetic field across $r=\infty$, which locally fixes the continuation without requiring any gluing prescription. In the subextremal $A>0$ sector considered here, $r=\infty$ is a timelike hypersurface connecting the exterior of one KBR region to the negative-$r$ interior of a neighboring one. Regularity of the metric and Maxwell field makes this continuation natural, and iterating it produces an infinite chain of KBR regions connected through wormhole-like bridges. The extension requires no thin shell and violates none of the standard energy conditions. It also exposes the ring singularity of the neighboring region without an intervening horizon, revealing a global structure fundamentally different from that suggested by a single KBR coordinate patch.

This global extension also changes the associated wave problem qualitatively. On the two-universe scattering region, the effective potential generically develops two barriers separated by a cavity across the gluing surface, and the corresponding QNM spectrum contains families of weakly damped modes suggestive of echoes. More importantly, we find two distinct manifestations of unstable modes. For $m=0$, we analytically establish the existence of purely imaginary exponentially growing modes for every $\ell$; the associated confining region coincides with the equatorial chronology-violating region containing closed timelike curves. In the $(\ell,m)=(2,2)$ sector, we find additional unstable branches for sufficiently large $a/M$ and small $BM$, driven by a black-hole-bomb mechanism in which the left potential barrier acts as a mirror, enabling repeated superradiant extraction from the right horizon. Because the scattering region is not globally hyperbolic, however, the existence of these growing modes does not by itself establish a conventional dynamical instability defined by evolution from generic Cauchy data.

From a quasi-local perspective, the outer horizon may be interpreted as a black-hole horizon. At the same time, the double-barrier potential supports weakly damped echo modes, a feature commonly associated with traversable wormhole geometries. Taken together, these results show that the natural extension across $r=\infty$ fundamentally changes the global interpretation of the KBR geometry and leads to a qualitatively distinct  spectrum.

The restriction to the subextremal $A>0$ sector is not essential to the extension itself. Beyond this restriction, the extension exhibits several qualitatively different causal structures. The extremal geometry remains extendible across a timelike $r=\infty$, whereas for $A=0$ this surface becomes null and for $A<0$ spacelike. A detailed treatment of these causal structures, together with the technical construction of the extension and the full spectral analysis, will be presented otherwhere separately.

We now turn to other possible extensions. The simplest, though less compelling, is to join distinct regular spacetime regions across $r=\infty$ by introducing matter shells. Another possibility is inspired by the $M=0$ case, where the KBR geometry reduces exactly to the Bertotti--Robinson (BR) spacetime, $\mathrm{AdS}_{2}\times S^{2}$, for which a transformation to global BR coordinates is known~\cite{Podolsky:2025tle}. Without this global description, the massless KBR chart can itself be repeatedly extended across $r=\infty$ using $y\sim1/r$, producing an infinite chain as in the main text. Global BR coordinates instead show that $r=\infty$ lies in the BR bulk, so these extensions merely pass between different massless KBR patches covered by a single global BR coordinate system, while the true null infinity becomes manifest only in the global coordinates. Nevertheless, for $M\neq0$, no analogous global coordinates are known. A natural place to seek a larger global description is the axial $\Omega=0$ locus, where the KBR coordinates degenerate. Introducing
\begin{eqnarray}
\varrho&=&\sqrt{\alpha^{2}(r-r_{\Omega})^{2}\cos^{2}\theta+(1+B^{2}r^{2})\sin^{2}\theta}\, ,\nonumber\\
\tan\Theta&=&\frac{\sqrt{1+B^{2}r^{2}}\sin\theta}{\alpha(r-r_{\Omega})\cos\theta}\, ,\nonumber\\
\tau&=&\frac{B\Delta_{\Omega}}{C\rho_{\Omega}^{2}\sqrt{1-A}}t\, ,\nonumber\\
\Phi&=&\varphi-\frac{a}{C\rho_{\Omega}^{2}}t\, ,
\end{eqnarray}
with $\alpha=B^{2}M\sqrt{I_{2}}/I_{1}$ and
$\rho_\Omega^2=r_\Omega^2+a^2$, locally resolves the two axial
$\Omega=0$ points, with $\varrho\to0$ corresponding to their blown-up
limit. Near $\varrho=0$, the metric takes, to leading order, the BR form
\begin{eqnarray}
\mathrm{d}s^{2}
\simeq \frac{C\rho_{\Omega}^{2}}{1+B^{2}r_\Omega^{2}}\left[
\frac{-\mathrm{d}\tau^{2}+\mathrm{d}\varrho^{2}}{\varrho^{2}}
+\mathrm{d}\Theta^{2}+\sin^{2}\Theta\,\mathrm{d}\Phi^{2}
\right]\,.\nonumber
\end{eqnarray}
Thus, if this coordinate transformation could be extended globally,
the $\Omega=0$ locus would correspond to a timelike BR-like null infinity. However, the coordinate system cannot be extended to a global one, since the Jacobian of the transformation is degenerate,
for example, at $\{r=r_\Omega,\theta=\pi/2\}$. Moreover, the nontrivial continuation of the Maxwell field provides an additional obstruction to a direct BR-like global description. The infinite chain we have constructed, or alternatively a period-two quotient, therefore provides a natural analytic extension, while other global descriptions remain open.

\section*{Acknowledgement}
This work is supported  by the National Key R\&D Program of China Grant No. 2022YFC2204603, and by the National Natural Science Foundation of China with grants No. 12475063,  No. 12247103, No. 12505067,  No. 12588101, and No. 12535002.

\bibliography{mainRef.bib}

@article{Cardoso:2004nk,
    author = "Cardoso, Vitor and Dias, Oscar J. C. and Lemos, Jose P. S. and Yoshida, Shijun",
    title = "{The Black hole bomb and superradiant instabilities}",
    eprint = "hep-th/0404096",
    archivePrefix = "arXiv",
    doi = "10.1103/PhysRevD.70.049903",
    journal = "Phys. Rev. D",
    volume = "70",
    pages = "044039",
    year = "2004",
    note = "[Erratum: Phys.Rev.D 70, 049903 (2004)]"
}

@article{Press:1972zz,
    author = "Press, William H. and Teukolsky, Saul A.",
    title = "{Floating Orbits, Superradiant Scattering and the Black-hole Bomb}",
    doi = "10.1038/238211a0",
    journal = "Nature",
    volume = "238",
    pages = "211--212",
    year = "1972"
}

@article{Singh:2026rbz,
    author = "Singh, Pradeep and Kala, Shubham and Nandan, Hemwati and Yousaf, M. and Atamurotov, Farruh and Mustafa, G.",
    title = "{Probing strong-gravity chaos in rotating Kerr{\textendash}Bertotti{\textendash}Robinson black holes}",
    doi = "10.1016/j.chaos.2026.118379",
    journal = "Chaos, Solitons {\&}amp; Fractals",
    volume = "208",
    pages = "118379",
    year = "2026"
}

@article{Ahmed:2026ozq,
    author = "Ahmed, Faizuddin and Donmez, Orhan and Al-Badawi, Ahmad and Sakalli, Izzet",
    title = "{Astrophysical signatures of Kerr-Bertotti-Robinson black holes in a cloud of strings: ISCO, microquasar QPOs, and Bondi-Hoyle-Lyttleton accretion}",
    eprint = "2605.10214",
    archivePrefix = "arXiv",
    primaryClass = "astro-ph.HE",
    month = "5",
    year = "2026"
}

@article{Huang:2026qzj,
    author = "Huang, Hai and Sun, Xudong and Chen, Juhua",
    title = "{Massless scalar scattering by Kerr-Bertotti-Robinson black holes:transparent-end channels and superradiance}",
    eprint = "2607.16068",
    archivePrefix = "arXiv",
    primaryClass = "gr-qc",
    month = "7",
    year = "2026"
}

@article{Hassanabadi:2026fzz,
    author = "Hassanabadi, Hassan and Good, Michael R. R. and Zare, Soroush and Luongo, Orlando and Kafikang, Fariba",
    title = "{Optical and thermodynamic properties of Kerr-Bertotti-Robinson black holes}",
    eprint = "2607.11979",
    archivePrefix = "arXiv",
    primaryClass = "gr-qc",
    month = "7",
    year = "2026"
}

@article{Roy:2026poj,
    author = "Roy, Saswati and Tapase, Anshul and Kala, Shubham and Nandan, Hemwati and Sen, Asoke Kumar",
    title = "{Light bending around the Kerr-Bertotti-Robinson black hole using material medium approach}",
    eprint = "2607.07437",
    archivePrefix = "arXiv",
    primaryClass = "gr-qc",
    month = "7",
    year = "2026"
}

@article{Wan:2026lca,
    author = "Wan, Xi and Zhang, Zhenyu and Wei, Fang-Stars and Hou, Yehui and Chen, Bin",
    title = "{Critical behavior of photon rings in Kerr-Bertotti-Robinson spacetime}",
    eprint = "2603.25049",
    archivePrefix = "arXiv",
    primaryClass = "gr-qc",
    doi = "10.1103/fyky-bkbg",
    journal = "Phys. Rev. D",
    volume = "114",
    number = "4",
    pages = "044005",
    year = "2026"
}

@article{Rehman:2026rzq,
    author = "Rehman, Hamza and Shaymatov, Sanjar and Hussain, Saddam and Zhu, Tao",
    title = "{Probing Kerr black hole in a uniform Bertotti-Robinson magnetic field through astrophysical quasi-periodic oscillations}",
    eprint = "2603.18129",
    archivePrefix = "arXiv",
    primaryClass = "astro-ph.HE",
    month = "3",
    year = "2026"
}

@article{Cardoso:2007az,
    author = "Cardoso, Vitor and Pani, Paolo and Cadoni, Mariano and Cavaglia, Marco",
    title = "{Ergoregion instability of ultracompact astrophysical objects}",
    eprint = "0709.0532",
    archivePrefix = "arXiv",
    primaryClass = "gr-qc",
    doi = "10.1103/PhysRevD.77.124044",
    journal = "Phys. Rev. D",
    volume = "77",
    pages = "124044",
    year = "2008"
}

@article{Comins,
 ISSN = {00804630},
 URL = {http://www.jstor.org/stable/79759},
 author = {N. Comins and B. F. Schutz},
 title = {On the Ergoregion Instability},
 journal = {Proceedings of the Royal Society of London. Series A, Mathematical and Physical Sciences},
 number = {1717},
 pages = {211--226},
 publisher = {The Royal Society},
 title = {On the Ergoregion Instability},
 urldate = {2026-08-14},
 volume = {364},
 year = {1978}
}

@article{Berti:2025hly,
    author = "Berti, Emanuele and others",
    editor = "Berti, Emanuele and Cardoso, Vitor and Carullo, Gregorio",
    title = "{Black hole spectroscopy: from theory to experiment}",
    eprint = "2505.23895",
    archivePrefix = "arXiv",
    primaryClass = "gr-qc",
    reportNumber = "RUP-25-10; YITP-25-64; RIKEN-iTHEMS-Report-25",
    doi = "10.1088/1361-6382/ae59e2",
    journal = "Class. Quant. Grav.",
    volume = "43",
    number = "12",
    pages = "123001",
    year = "2026"
}

@article{Ashtekar:1998sp,
    author = "Ashtekar, Abhay and Beetle, Christopher and Fairhurst, Stephen",
    title = "{Isolated horizons: A Generalization of black hole mechanics}",
    eprint = "gr-qc/9812065",
    archivePrefix = "arXiv",
    doi = "10.1088/0264-9381/16/2/027",
    journal = "Class. Quant. Grav.",
    volume = "16",
    pages = "L1--L7",
    year = "1999"
}

@article{Ashtekar:2004cn,
    author = "Ashtekar, Abhay and Krishnan, Badri",
    title = "{Isolated and dynamical horizons and their applications}",
    eprint = "gr-qc/0407042",
    archivePrefix = "arXiv",
    doi = "10.12942/lrr-2004-10",
    journal = "Living Rev. Rel.",
    volume = "7",
    pages = "10",
    year = "2004"
}

@article{Dotti:2011ix,
    author = "Dotti, Gustavo and Gleiser, Reinaldo J. and Ranea-Sandoval, Ignacio F.",
    title = "{Unstable fields in Kerr spacetimes}",
    eprint = "1111.5974",
    archivePrefix = "arXiv",
    primaryClass = "gr-qc",
    doi = "10.1088/0264-9381/29/9/095017",
    journal = "Class. Quant. Grav.",
    volume = "29",
    pages = "095017",
    year = "2012"
}

@article{Ahmed:2025ril,
    archiveprefix = {arXiv},
    author        = {Ahmed, Faizuddin and Sakall{\i}, {\.I}zzet and Al-Badawi, Ahmad},
    eprint        = {2511.11792},
    month         = {11},
    primaryclass  = {gr-qc},
    title         = {{Kerr-Bertotti-Robinson Black Holes Surrounded by a Cloud of Strings}},
    year          = {2025}
}

@article{Ali:2025beh,
    archiveprefix = {arXiv},
    author        = {Ali, Heena and Ghosh, Sushant G.},
    doi           = {10.1088/1475-7516/2026/01/018},
    eprint        = {2508.15862},
    journal       = {JCAP},
    pages         = {018},
    primaryclass  = {gr-qc},
    title         = {{Parameter estimation of Kerr-Bertotti-Robinson black holes using their shadows}},
    volume        = {01},
    year          = {2026}
}

@article{Astorino:2025lih,
    archiveprefix = {arXiv},
    author        = {Astorino, Marco},
    doi           = {10.1103/c5lw-53yd},
    eprint        = {2508.12908},
    journal       = {Phys. Rev. D},
    number        = {10},
    pages         = {104077},
    primaryclass  = {gr-qc},
    reportnumber  = {LIFT-11-4.25},
    title         = {{Black holes in the external Bertotti-Robinson-Bonnor-Melvin electromagnetic field}},
    volume        = {112},
    year          = {2025}
}

@article{Berti:2009kk,
    archiveprefix = {arXiv},
    author        = {Berti, Emanuele and Cardoso, Vitor and Starinets, Andrei O.},
    doi           = {10.1088/0264-9381/26/16/163001},
    eprint        = {0905.2975},
    journal       = {Class. Quant. Grav.},
    pages         = {163001},
    primaryclass  = {gr-qc},
    title         = {{Quasinormal modes of black holes and black branes}},
    volume        = {26},
    year          = {2009}
}

@article{Chen:2024rov,
    archiveprefix = {arXiv},
    author        = {Chen, Tan and Cai, Rong-Gen and Hu, Bin},
    doi           = {10.1103/PhysRevD.109.084049},
    eprint        = {2402.02027},
    journal       = {Phys. Rev. D},
    number        = {8},
    pages         = {084049},
    primaryclass  = {gr-qc},
    title         = {{Quasinormal modes of gravitational perturbation for uniformly accelerated black holes}},
    volume        = {109},
    year          = {2024}
}

@article{Destounis:2020pjk,
    archiveprefix = {arXiv},
    author        = {Destounis, Kyriakos and Fontana, Rodrigo D. B. and Mena, Filipe C.},
    doi           = {10.1103/PhysRevD.102.044005},
    eprint        = {2005.03028},
    journal       = {Phys. Rev. D},
    number        = {4},
    pages         = {044005},
    primaryclass  = {gr-qc},
    title         = {{Accelerating black holes: quasinormal modes and late-time tails}},
    volume        = {102},
    year          = {2020}
}

@article{Fiziev:2011mm,
    archiveprefix = {arXiv},
    author        = {Fiziev, Plamen and Staicova, Denitsa},
    doi           = {10.1103/PhysRevD.84.127502},
    eprint        = {1109.1532},
    journal       = {Phys. Rev. D},
    pages         = {127502},
    primaryclass  = {gr-qc},
    reportnumber  = {SU-TH-7-09-2011},
    title         = {{Application of the confluent Heun functions for finding the quasinormal modes of nonrotating black holes}},
    volume        = {84},
    year          = {2011}
}

@article{Gray:2025lwy,
    archiveprefix = {arXiv},
    author        = {Gray, Finnian and Kubiznak, David and Ovcharenko, Hryhorii and Podolsky, Jiri},
    doi           = {10.1103/8832-htpg},
    eprint        = {2511.21538},
    journal       = {Phys. Rev. D},
    number        = {4},
    pages         = {044050},
    primaryclass  = {gr-qc},
    title         = {{Hidden symmetries and separability structures of Ovcharenko-Podolsk{\'y} and conformal-to-Carter spacetimes}},
    volume        = {113},
    year          = {2026}
}

@article{Hatsuda:2020sbn,
    archiveprefix = {arXiv},
    author        = {Hatsuda, Yasuyuki},
    doi           = {10.1088/1361-6382/abc82e},
    eprint        = {2006.08957},
    journal       = {Class. Quant. Grav.},
    number        = {2},
    pages         = {025015},
    primaryclass  = {gr-qc},
    reportnumber  = {RUP-20-19},
    title         = {{Quasinormal modes of Kerr-de Sitter black holes via the Heun function}},
    volume        = {38},
    year          = {2020}
}

@article{Hu:2026slp,
    archiveprefix = {arXiv},
    author        = {Hu, Li and Cai, Rong-Gen and Wang, Shao-Jiang},
    eprint        = {2603.18821},
    month         = {3},
    primaryclass  = {gr-qc},
    title         = {{Thermodynamics of Kerr-Bertotti-Robinson black hole}},
    year          = {2026}
}

@article{Israel:1966rt,
    author  = {Israel, W.},
    doi     = {10.1007/BF02710419},
    journal = {Nuovo Cim. B},
    note    = {[Erratum: Nuovo Cim.B 48, 463 (1967)]},
    pages   = {1},
    title   = {{Singular hypersurfaces and thin shells in general relativity}},
    volume  = {44S10},
    year    = {1966}
}

@article{Kokkotas:1999bd,
    archiveprefix = {arXiv},
    author        = {Kokkotas, Kostas D. and Schmidt, Bernd G.},
    doi           = {10.12942/lrr-1999-2},
    eprint        = {gr-qc/9909058},
    journal       = {Living Rev. Rel.},
    pages         = {2},
    title         = {{Quasinormal modes of stars and black holes}},
    volume        = {2},
    year          = {1999}
}

@article{Konoplya:2011qq,
    archiveprefix = {arXiv},
    author        = {Konoplya, R. A. and Zhidenko, A.},
    doi           = {10.1103/RevModPhys.83.793},
    eprint        = {1102.4014},
    journal       = {Rev. Mod. Phys.},
    pages         = {793--836},
    primaryclass  = {gr-qc},
    title         = {{Quasinormal modes of black holes: From astrophysics to string theory}},
    volume        = {83},
    year          = {2011}
}

@article{Li:2025rtf,
    archiveprefix = {arXiv},
    author        = {Li, Xiang-Qian and Yan, Hao-Peng and Yue, Xiao-Jun},
    doi           = {10.1140/epjc/s10052-026-15441-5},
    eprint        = {2512.02921},
    journal       = {Eur. Phys. J. C},
    number        = {2},
    pages         = {176},
    primaryclass  = {gr-qc},
    title         = {{Gravitational-wave imprints of Kerr{\textendash}Bertotti{\textendash}Robinson black holes: frequency blue-shift and waveform dephasing}},
    volume        = {86},
    year          = {2026}
}

@article{Liu:2025wwq,
    author = "Liu, Wentao and Liu, Yang and Wu, Di and Liu, Yu-Xiao",
    title = "{Universal framework for horizon-scale tests of gravity with black hole shadows}",
    eprint = "2511.06017",
    archivePrefix = "arXiv",
    primaryClass = "gr-qc",
    doi = "10.1103/66x9-zj16",
    journal = "Phys. Rev. D",
    volume = "114",
    number = "2",
    pages = "L021503",
    year = "2026"
}

@article{Lu:2026kcm,
    archiveprefix = {arXiv},
    author        = {Lu, Junjie and Wu, Xin},
    doi           = {10.1140/epjc/s10052-026-15482-w},
    eprint        = {2603.12674},
    journal       = {Eur. Phys. J. C},
    number        = {3},
    pages         = {256},
    primaryclass  = {gr-qc},
    title         = {{Third type of spacetime with the coexistence of integrability and non-integrability}},
    volume        = {86},
    year          = {2026}
}

@article{Mars:1993mj,
    archiveprefix = {arXiv},
    author        = {Mars, Marc and Senovilla, Jose M. M.},
    doi           = {10.1088/0264-9381/10/9/026},
    eprint        = {gr-qc/0201054},
    journal       = {Class. Quant. Grav.},
    pages         = {1865--1897},
    title         = {{Geometry of general hypersurfaces in space-time: Junction conditions}},
    volume        = {10},
    year          = {1993}
}

@article{Mirkhaydarov:2026fyn,
    archiveprefix = {arXiv},
    author        = {Mirkhaydarov, Mirjavoxir and Xamidov, Tursunali and Sheoran, Pankaj and Shaymatov, Sanjar and Nandan, Hemwati},
    eprint        = {2601.09919},
    month         = {1},
    primaryclass  = {gr-qc},
    title         = {{Non-Monotonic Enhancement of the Magnetic Penrose Process in Kerr-Bertotti-Robinson Spacetime and its Implication for Electron Acceleration}},
    year          = {2026}
}

@article{Motohashi:2021zyv,
    archiveprefix = {arXiv},
    author        = {Motohashi, Hayato and Noda, Sousuke},
    doi           = {10.1093/ptep/ptac020},
    eprint        = {2103.10802},
    journal       = {PTEP},
    number        = {8},
    pages         = {083E03},
    primaryclass  = {gr-qc},
    title         = {{Exact solution for wave scattering from black holes: Formulation}},
    volume        = {2021},
    year          = {2021}
}

@article{Mustafa:2026gly,
    author = "Mustafa, G. and Donmez, Orhan and Gogoi, Dhruba Jyoti and Ghosh, Sushant G. and Hussain, Ibrar and Yuan, Chengxun",
    title = "{Dynamics, ringdown, and accretion-driven multiple quasi-periodic oscillations of Kerr{\textendash}Bertotti{\textendash}Robinson black holes}",
    eprint = "2602.08911",
    archivePrefix = "arXiv",
    primaryClass = "gr-qc",
    doi = "10.1088/1475-7516/2026/07/065",
    journal = "JCAP",
    volume = "07",
    pages = "065",
    year = "2026"
}

@article{Nollert:1999ji,
    author  = {Nollert, Hans-Peter},
    doi     = {10.1088/0264-9381/16/12/201},
    journal = {Class. Quant. Grav.},
    pages   = {R159--R216},
    title   = {{TOPICAL REVIEW: Quasinormal modes: the characteristic `sound' of black holes and neutron stars}},
    volume  = {16},
    year    = {1999}
}

@article{Ovcharenko:2025cpm,
    archiveprefix = {arXiv},
    author        = {Ovcharenko, Hryhorii and Podolsk{\'y}, Ji{\v{r}}{\'\i}},
    doi           = {10.1103/8wkz-th6v},
    eprint        = {2508.04850},
    journal       = {Phys. Rev. D},
    number        = {6},
    pages         = {064076},
    primaryclass  = {gr-qc},
    title         = {{New class of rotating charged black holes with nonaligned electromagnetic field}},
    volume        = {112},
    year          = {2025}
}

@article{Patrick:2018orp,
    archiveprefix = {arXiv},
    author        = {Patrick, Sam and Coutant, Antonin and Richartz, Maur{\'\i}cio and Weinfurtner, Silke},
    doi           = {10.1103/PhysRevLett.121.061101},
    eprint        = {1801.08473},
    journal       = {Phys. Rev. Lett.},
    number        = {6},
    pages         = {061101},
    primaryclass  = {gr-qc},
    title         = {{Black hole quasibound states from a draining bathtub vortex flow}},
    volume        = {121},
    year          = {2018}
}

@article{Podolsky:2025tle,
    archiveprefix = {arXiv},
    author        = {Podolsky, Jiri and Ovcharenko, Hryhorii},
    doi           = {10.1103/rfgv-ybz5},
    eprint        = {2507.05199},
    journal       = {Phys. Rev. Lett.},
    number        = {18},
    pages         = {181401},
    primaryclass  = {gr-qc},
    title         = {{Kerr Black Hole in a Uniform Bertotti-Robinson Magnetic Field: An Exact Solution}},
    volume        = {135},
    year          = {2025}
}

@article{Siahaan:2025ngu,
    archiveprefix = {arXiv},
    author        = {Siahaan, Haryanto M.},
    eprint        = {2512.12533},
    month         = {12},
    primaryclass  = {gr-qc},
    title         = {{Kerr-Bertotti-Robinson Spacetime and the Kerr/CFT Correspondence}},
    year          = {2025}
}

@article{Siahaan:2026tuf,
    archiveprefix = {arXiv},
    author        = {Siahaan, Haryanto M.},
    eprint        = {2603.00653},
    month         = {2},
    primaryclass  = {gr-qc},
    title         = {{Meissner Effect in Kerr--Bertotti--Robinson Spacetime}},
    year          = {2026}
}

@article{Suzuki:1998vy,
    archiveprefix = {arXiv},
    author        = {Suzuki, Hisao and Takasugi, Eiichi and Umetsu, Hiroshi},
    doi           = {10.1143/PTP.100.491},
    eprint        = {gr-qc/9805064},
    journal       = {Prog. Theor. Phys.},
    pages         = {491--505},
    reportnumber  = {EPHOU-98-005, OU-HET-296},
    title         = {{Perturbations of Kerr-de Sitter black hole and Heun's equations}},
    volume        = {100},
    year          = {1998}
}

@inproceedings{Wald:1997wa,
    archiveprefix = {arXiv},
    author        = {Wald, Robert M.},
    doi           = {10.1007/978-94-017-0934-7_5},
    eprint        = {gr-qc/9710068},
    month         = {10},
    pages         = {69--85},
    reportnumber  = {EFI-97-43},
    title         = {{Gravitational collapse and cosmic censorship}},
    year          = {1997}
}

@article{Wang:2025bjf,
    archiveprefix = {arXiv},
    author        = {Wang, Tower},
    eprint        = {2508.04684},
    month         = {8},
    primaryclass  = {gr-qc},
    title         = {{Innermost stable circular orbit of Kerr-Bertotti-Robinson black holes and inspirals from it: Exact solutions}},
    year          = {2025}
}

@article{Wang:2025vsx,
    archiveprefix = {arXiv},
    author        = {Wang, Xinyu and Hou, Yehui and Wan, Xi and Guo, Minyong and Chen, Bin},
    doi           = {10.1088/1475-7516/2026/02/050},
    eprint        = {2507.22494},
    journal       = {JCAP},
    pages         = {050},
    primaryclass  = {gr-qc},
    title         = {{Geodesics and shadows in the Kerr-Bertotti-Robinson black hole spacetime}},
    volume        = {02},
    year          = {2026}
}

@article{Wang:2026czl,
    archiveprefix = {arXiv},
    author        = {Wang, Chao-Hui and Meng, Xiang-Cheng and Wei, Shao-Wen},
    eprint        = {2602.03161},
    month         = {2},
    primaryclass  = {gr-qc},
    title         = {{Magnetic field effects on spherical orbit in Kerr-Bertotti-Robinson spacetime: constraints from jet precession of M87*}},
    year          = {2026}
}

@article{Wu:2025wbp,
    archiveprefix = {arXiv},
    author        = {Wu, Liang-Bi and Xie, Libo and Cao, Li-Ming and Ji, Ming-Fei and Zhou, Yu-Sen},
    doi           = {10.1007/s11433-025-2912-y},
    eprint        = {2512.06903},
    journal       = {Sci. China Phys. Mech. Astron.},
    number        = {4},
    pages         = {240415},
    primaryclass  = {gr-qc},
    title         = {{Quasinormal modes of Schwarzschild-de Sitter black holes in semi-open systems}},
    volume        = {69},
    year          = {2026}
}

@article{Xiong:2023usm,
    archiveprefix = {arXiv},
    author        = {Xiong, Wei and Li, Peng-Cheng},
    doi           = {10.1103/PhysRevD.108.044064},
    eprint        = {2305.04040},
    journal       = {Phys. Rev. D},
    number        = {4},
    pages         = {044064},
    primaryclass  = {gr-qc},
    title         = {{Quasinormal modes of rotating accelerating black holes}},
    volume        = {108},
    year          = {2023}
}

@article{Zeng:2025olq,
    archiveprefix = {arXiv},
    author        = {Zeng, Xiao-Xiong and Wang, Ke},
    doi           = {10.1103/vc96-snjm},
    eprint        = {2507.21777},
    journal       = {Phys. Rev. D},
    number        = {6},
    pages         = {064032},
    primaryclass  = {gr-qc},
    title         = {{Energy extraction from the Kerr-Bertotti-Robinson black hole via magnetic reconnection in a circular and a plunging plasma}},
    volume        = {112},
    year          = {2025}
}

@article{Zeng:2025tji,
    archiveprefix = {arXiv},
    author        = {Zeng, Xiao-Xiong and Yang, Chen-Yu and Yu, Hao},
    doi           = {10.1140/epjc/s10052-025-14989-y},
    eprint        = {2508.03020},
    journal       = {Eur. Phys. J. C},
    number        = {11},
    pages         = {1242},
    primaryclass  = {gr-qc},
    title         = {{Optical characteristics of the Kerr{\textendash}Bertotti{\textendash}Robinson black hole}},
    volume        = {85},
    year          = {2025}
}

@article{Zhang:2025ole,
    archiveprefix = {arXiv},
    author        = {Zhang, Yu-Kun and Wei, Shao-Wen},
    doi           = {10.1103/dmh3-ht32},
    eprint        = {2510.07914},
    journal       = {Phys. Rev. D},
    number        = {10},
    pages         = {104024},
    primaryclass  = {gr-qc},
    title         = {{Effects of magnetic fields on spinning test particles orbiting Kerr-Bertotti-Robinson black holes}},
    volume        = {113},
    year          = {2026}
}
\bibliographystyle{apsrev4-1}

\end{document}